\RequirePackage[2020-02-02]{latexrelease}
\documentclass[superscriptaddress,secnumarabic,
amssymb,amsmath,nobibnotes,aps,prd,showkeys,showpacs,nofootinbib]{revtex4}%
\usepackage{graphicx}
\usepackage{epsf}
\usepackage{bm}
\usepackage{amsmath}
\usepackage{amsthm}
\usepackage{amsfonts}
\usepackage{amssymb}
\usepackage{epstopdf}
\usepackage{hhline}
\usepackage{tabularx}
\usepackage{multirow}
\usepackage{booktabs}
\usepackage{wrapfig}
\usepackage{comment}

\usepackage{lipsum}
\usepackage{natbib}%
\providecommand{\U}[1]{\protect\rule{.1in}{.1in}}
\newcommand{\be}{\begin{equation}}
\newcommand{\ee}{\end{equation}}
\usepackage{color}

\begin{document}

\title{Complexity factor Parametrization for Traversable Wormholes}

\author{Subhra Bhattacharya}
\email{subhra.maths@presiuniv.ac.in}
\author{Subhasis Nalui}
\email{subhasis222nalui@gmail.com}
\affiliation{Department of Mathematics, Presidency University, Kolkata-700073, India}
\begin{abstract}
It is known that static traversable wormhole in Einstein gravity is supported by matter that violates null energy conditions (NEC). Essentially such wormhole will be characterised by a central throat with anisotropic matter lining the throat that violates NEC. This in turn provides viable geometry for the wormhole to sustain. In 2018, L. Herrera introduced a new classification for spherically symmetric bodies called ``complexity factor". It was proposed that a spherically symmetric non trivial geometry can be classified as complex or non-complex based on the nature of the inhomogeneity and anisotropy of the stress energy tensors with only homogeneous and isotropic matter distribution leading to null complexity. Mathematically there was also another way of obtaining zero complexity geometry. In this context since static traversable wormhole by default is characterised by anisotropic and inhomogeneous matter stress tensors, the question we answer is whether it is possible to obtain zero complexity class of wormholes supported by exotic matter.

\end{abstract}

\maketitle
\section{Introduction}
\label{Intro}

In 2018 Herera \cite{1} defined ``complexity factor" corresponding to the modifications introduced in the gravitational mass of a homogeneous, self gravitating, static and spherically symmetric system, due to the inhomogeneity and anisotropy of matter stress energy tensors. The complexity factor was obtained by the orthogonal splitting of the Riemann curvature tensor \cite{os} and hence contained information on the internal structure of the self gravitating body related to its fluid distribution. Objects with zero complexity were classified as the simplest systems that agreed to a homogeneous and isotropic distribution of matter. But simple systems were not just restricted to the class of homogeneous and isotropic matter distributions. Systems could be simple even in the presence of inhomogeneity and anisotropy of fluid if the two factors contributing to the complexity mutually cancelled each other. This gave an equivalence class of systems, classified based on their similar complexity. The complexity factor has recently been applied in a variety of gravitating objects and scenarios, as in the case of anisotropic star models \cite{2} and the gravitational cracking scenario \cite{3}. Zero complexity configurations were obtained in \cite{4} using gravitational decoupling of static spherically symmetric systems. It has been recently used in the context of wormholes \cite{5}. The definition of complexity factor has also been extended to dynamical spherically symmetric dissipative systems \cite{6}. Other work using complexity factor is listed in \cite{7}.

Wormholes, are hypothetical structures in the Universe, whose existence is yet to be confirmed. Despite being speculative, these gravitational objects have generated ceaseless excitement in the scientific community. Existence of structures resembling, what we call now wormhole, was known since 1916 \cite{flamm}. Later terms like ``drainhole", ``kinks" and ``tunnels" were used to describe similar structures \cite{8}.   Now several different solutions and variations of wormhole exists \cite{book}. Continued interest on wormhole has resulted in the growth of a vibrant literature on them. Yet they are unusual and their existence is prohibited by the laws of physics. Hence special techniques were adopted to obtain their solutions. In \cite{mt} Lorentzian wormholes were obtained by the reconstruction procedure, where the space time geometry was determined after imposing viable bridge conditions corresponding to two asymptotically flat space time. Such a wormhole required exotic matter tensors that violated null energy conditions (NEC). Although this in itself might seem non-physical, yet there are known processes in the early universe that could have created such tiny pockets of exotic matter where it might be possible to detect their existence \cite{9}. In \cite{10} ways of actually observing wormholes using astrophysical observations is discussed. Some recent articles discusses possible detection of wormholes in the active galactic nuclei \cite{11} and in the bulge of the milky way galaxy \cite{12}. Recently traversable wormholes were constructed in Einstein-Dirac-Maxwell theory without exotic matter \cite{13}. This led to a flurry of discussions on the existence of such wormholes in nature \cite{14}. Finally in \cite{15} it was shown that wormholes that are asymmetric with respect to the throat could be supported by normalizable and smooth matter. This work was further analysed in \cite{16} to study the quasinormal modes, echoes and radius of shadow for the corresponding wormhole configurations. 

It is evident that viable wormhole solution within the general relativistic framework is an important aspect of modern astrophysics. In this context, the complexity definition for spherically symmetric gravitational objects adds another tool to obtain wormhole geometries that are equivalent to less complex gravitational structures. In order to determine wormhole structure with suitable geometry as well as matter one needs to work out five unknowns from the three Einstein's field equations and one fluid conservation equation. Since the equation for complexity factor adds another constraint to the set of equations, our aim is to use this to generate suitable simple structure corresponding to a wormhole. It would be also interesting to comment on the existence of ``zero-complexity" or ``simple" wormholes. It is known that zero-complexity systems can be obtained in the presence of homogeneous and isotropic matter or by mathematical cancellation of terms. Therefore we will also try to answer whether some wormholes can be classified into the zero-complexity class, given that they are essentially supported by inhomogeneous matter.

The paper is organised as follows: In section 2 we give a general discussion on the complexity factor with respect to (w.r.t) a traversable wormhole. In section 3 we provide the master equations for evaluating a traversable wormhole using the complexity factor and provide ways of solving them with examples. In section 4 we discuss the general zero-complexity wormhole. We end with a short conclusion in section 5.

\section{Complexity Factor in the context of the Wormhole}

We consider the general static spherically symmetric metric \cite{mt} that can support a wormhole geometry:

\begin{equation}
ds^{2}=-e^{2\phi(r)}dt^{2}+\frac{dr^{2}}{1-\frac{b(r)}{r}}+r^{2}(d\theta^{2}+\sin^{2}\theta d\Phi^{2})\label{metric}
\end{equation}

It was discussed in \cite{mt} that the above metric can harbour a wormhole provided viable restrictions were imposed on the metric. Here the radial coordinate $r$ varies between $r_{0}\leq r< +\infty$ with $r_{0}$ identified as the throat of the wormhole whose shape is defined using the function $b(r).$ In order to have continuity over the entire space time a similar space time would exist as $r\rightarrow -\infty.$ At the throat the shape function $b(r)$ must satisfy $b(r_{0})=r_{0}$ with $b'(r_{0})<1.$ Further for asymptotically flat wormhole geometry the shape function needs to satisfy $\frac{b(r)}{r}\rightarrow 0$ as $r\rightarrow\infty.$ Wormholes will be traversable only if there are no event horizons, which is ensured by a finite $\phi(r)$ for all $r.$ With the components of the inhomogeneous and anisotropic matter stress energy tensor $T_{\mu\nu}$ defined by $(\rho,p_{r},p_{t})$ the Einstein's field equations $G_{\mu\nu}=\kappa T_{\mu\nu}$ with $\kappa=8\pi G=c^{2}=1,$ give:
\begin{align}
\rho(r)=&\frac{b'}{r^{2}}\label{rho}\\
 p_{r}(r)=&-\frac{b}{r^{3}}+2\left(1-\frac{b}{r}\right)\frac{\phi'}{r}\label{pr}\\
 p_{t}(r)=&\left(1-\frac{b}{r}\right)\left[\phi''+(\phi')^{2}+\frac{\phi'}{r}-\frac{b'r-b}{2r(r-b)}\phi'-\frac{b'r-b}{2r^{2}(r-b)}\right]\label{pt}
\end{align} 
with the energy momentum tensor satisfying the conservation equations $T^{\mu}_{\nu;\mu}$=0 as given by:
\begin{equation}
p'_{r}=\frac{2}{r}(p_{t}-p_{r})-(\rho+p_{r})\phi'(r).\label{consv}
\end{equation} 
The complexity factor corresponding to a spherically symmetric space time is defined as \cite{1}
\begin{equation}
y_{tf}=p_{r}-p_{t}-\frac{1}{2r^{3}}\int_{0}^{r}s^{3}\rho'(s)ds\label{cf}
\end{equation}
The equation shows that for a homogeneous and isotropic distribution of matter, $y_{tf}$ is zero. Further if for certain matter distribution $(p_{r}-p_{t})=\frac{1}{2r^{3}}\int_{0}^{r}s^{3}\rho'(s)ds$ then again, the complexity factor is zero. Therefore it is evident that simple systems can exist in inhomogeneous and anisotropic distribution of matter. Wormholes, by definition can be supported by only inhomogeneous distribution of matter. So zero complexity wormhole can only be obtained in the second case. It may be noted that the complexity factor for a wormhole is suitably defined as $y_{tf}=(p_{r}-p_{t})-\frac{1}{2r^{3}}\int_{r_{0}}^{r}s^{3}\rho'(s)ds$ \cite{5}. From the expression, it is evident that $y_{tf}$ carries all the information regarding the geometric properties of the wormhole. Our aim is to use this information to generate viable wormhole geometries. Further we would like to address whether it is possible to define a zero complexity wormhole solutions. 

\section{General Wormhole representation using the Complexity Factor}

Using the field equation (\ref{rho}-\ref{pt}) in equation (\ref{cf}) the complexity factor for the wormhole can be written as:
\begin{equation}
y_{tf}=\left(1-\frac{b(r)}{r}\right)\left[\frac{\phi'}{r}-(\phi')^{2}-\phi''\right]+\phi'\left(\frac{rb'(r)-b(r)}{2r^{2}}\right)+\frac{C}{2r^{3}}\label{wcf}
\end{equation}
with $C=r_{0}b'(r_{0})-3r_{0},$ constant. In the above equation there are three unknowns, the shape function $b(r),$ the red-shift factor $\phi(r)$ and the complexity factor $y_{tf}.$ Evidently knowledge of at least two is required to determine the third. Accordingly we rewrite the above equation in $\frac{b(r)}{r}$ as:
\begin{equation}
\left(\frac{b(r)}{r}\right)'-\left(\frac{b(r)}{r}\right)\left(\frac{2}{r}-2\phi'-\frac{2\phi''}{\phi'}\right)=\frac{2}{\phi'}\left(y_{tf}-\frac{C}{2r^{3}}\right)-\left(\frac{2}{r}-2\phi'-\frac{2\phi''}{\phi'}\right)
\end{equation}
This being a linear ordinary differential equation, $\frac{b(r)}{r}$ can be solved as:
\begin{equation}
\frac{b(r)}{r}=1+\left[2\int\frac{e^{2\phi}\phi'}{r^{2}}\left(y_{tf}-\frac{C}{2r^{3}}\right)dr+D\right]\left(\frac{r}{\phi'}\right)^{2}e^{-2\phi}\label{me}
\end{equation}
where $D$ is the constant of integration.
For (\ref{me}) to be a traversable wormhole throat, it must satisfy the following conditions:
\begin{itemize}
\item[a)] As $r\rightarrow\infty,$ both $\frac{b(r)}{r}$ and $\phi(r)$ should be bounded, $\Rightarrow~D=0$ and that $2\left(\frac{r}{\phi'}\right)^{2}e^{-2\phi}\int\frac{e^{2\phi}\phi'}{r^{2}}\left(y_{tf}-\frac{C}{2r^{3}}\right)dr\rightarrow -1.$
\item[b)] At the wormhole throat $r=r_{0}$ we have $\frac{b(r)}{r}=1$ which gives $2\int\frac{e^{2\phi}\phi'}{r^{2}}\left(y_{tf}-\frac{C}{2r^{3}}\right)dr\mid_{r=r_{0}}=0.$
\item[c)] Also it must satisfy the compatibility condition that $b'(r_{0})=1+2\frac{1+r_{0}^{2}y_{tf0}}{1+r_{0}\phi'(r_{0})}$ where $y_{tf0}$ is $y_{tf}$ at $r=r_{0}.$ 
\end{itemize}

This shows that corresponding to any general complexity function $y_{tf}$ we can get a traversable wormhole from equation (\ref{me}), provided it is solvable. Here we sketch a simple method of solving the integral in (\ref{me}). 

\subsection{A Polynomial Approach}

We start with the corresponding polynomial representation for $y_{tf}$
\begin{equation}
y_{tf}=a_{0}+a_{1}\left(\frac{r_{0}}{r}\right)+a_{2}\left(\frac{r_{0}}{r}\right)^{2}+a_{3}\left(\frac{r_{0}}{r}\right)^{3}+a_{4}\left(\frac{r_{0}}{r}\right)^{4}+\cdots+a_{n}\left(\frac{r_{0}}{r}\right)^{n}+\frac{C}{2r^{3}}\label{ytf1}
\end{equation} 
with $C$ as defined above, (this last term is taken to simplify calculations) and $a_{0},a_{1},\ldots,a_{n}$ are all arbitrary constants for $n\in\mathbb{Z+}.$ The red-shift function is assumed as $e^{2\phi(r)}=1+\left(\frac{r}{r_{0}}\right)^{2\alpha},~\alpha<0.$ Using (\ref{ytf1}) in (\ref{me}) we get:
\begin{equation}
\begin{split}
\frac{b(r)}{r}=1+\frac{2r_{0}^{2}}{\alpha}\left(1+\left(\frac{r}{r_{0}}\right)^{2\alpha}\right)\left[\frac{a_{0}}{2\alpha-2}\left(\frac{r}{r_{0}}\right)^{2-2\alpha}+\frac{a_{1}}{2\alpha-3}\left(\frac{r}{r_{0}}\right)^{1-2\alpha}+\frac{a_{2}}{2\alpha-4}\left(\frac{r}{r_{0}}\right)^{-2\alpha}\right.\\
\left.+\frac{a_{3}}{2\alpha-5}\left(\frac{r}{r_{0}}\right)^{-1-2\alpha}+\frac{a_{4}}{2\alpha-6}\left(\frac{r}{r_{0}}\right)^{-2-2\alpha}+\cdots+\frac{a_{n}}{2\alpha-n-2}\left(\frac{r}{r_{0}}\right)^{2-n-2\alpha}\right]\label{shpn}
\end{split}
\end{equation}
 The above function $b(r)$ will describe a suitable traversable wormhole provided it fulfils the constraint conditions (a)-(c). Using (a) we get $a_{0}=a_{1}=a_{2}=0.$ Further analysing the coefficient of $a_{3}$ we get $-\frac{1}{2}\leq \alpha\leq 0.$ Thus (\ref{shpn}) for $n\geq 3$ gives us a wormhole throat with complexity $y_{tf}= a_{3}\left(\frac{r_{0}}{r}\right)^{3}+a_{4}\left(\frac{r_{0}}{r}\right)^{4}+\cdots+a_{n}\left(\frac{r_{0}}{r}\right)^{n}+\frac{C}{2r^{3}}.$

 \subsubsection{No wormhole for $n=3$}
 
Substituting $n=3$ gives $y_{tf}=a_{3}\left(\frac{r_{0}}{r}\right)^{3}+\frac{C}{2r^{3}}$ and $\frac{b(r)}{r}=1+\frac{2r_{0}^{2}a_{3}}{\alpha(2\alpha-5)}\left[1+\left(\frac{r}{r_{0}}\right)^{2\alpha}\right]\left(\frac{r}{r_{0}}\right)^{-1-2\alpha}.$ Simple calculations show that this shape function does not satisfy the constraints for a wormhole. Accordingly for a polynomial $y_{tf}$ with the given red-shift function $n=3$ does not give us any viable wormhole.
 
\subsubsection{Wormhole for $n=4$} 

For $n=4$ we have 
\begin{align*}
y_{tf}&=+a_{3}\left(\frac{r_{0}}{r}\right)^{3}+a_{4}\left(\frac{r_{0}}{r}\right)^{4}+\frac{C}{2r^{3}}\\
\frac{b(r)}{r}&=1+\frac{2r_{0}^{2}}{\alpha}\left[1+\left(\frac{r}{r_{0}}\right)^{2\alpha}\right]\left(\frac{r}{r_{0}}\right)^{-1-2\alpha}\left[\frac{a_{3}}{2\alpha-5}+\frac{a_{4}}{2\alpha-6}\left(\frac{r}{r_{0}}\right)^{-1}\right]
\end{align*}
Imposing the constraints (a)-(c) gives us i) $\alpha=-\frac{1}{2},$ ii) $a_{3}=-\frac{3}{2r_{0}^{2}}$ and iii) $a_{4}=\frac{7}{4r_{0}^{2}}.$ The corresponding wormhole for $n=4$ is given by:
\begin{equation*}
\frac{b(r)}{r}=\left(\frac{r_{0}}{r}\right)^{2},\qquad
e^{2\phi(r)}=1+\left(\frac{r_{0}}{r}\right)\qquad\text{and}\qquad
y_{tf}=\frac{7r_{0}}{2r^{3}}\left[-1+\frac{1}{2}\left(\frac{r_{0}}{r}\right)\right]
\end{equation*}

\subsubsection{Wormhole for $n=5$}

A similar calculation for $n=5$ and $\alpha=-\frac{1}{2}$ gives the following wormhole:
\begin{align*}
\frac{b(r)}{r}=&\left(\frac{r_{0}}{r}\right)^{2}-\frac{a_{5}r_{0}^{2}}{2}\left[\left(\frac{r_{0}}{r}\right)-\left(\frac{r_{0}}{r}\right)^{3}\right]\\
e^{2\phi(r)}=&1+\left(\frac{r_{0}}{r}\right)\\
y_{tf}=&-\left(\frac{7}{2r_{0}^{2}}+\frac{a_{5}}{2}\right)\left(\frac{r_{0}}{r}\right)^{3}+\left(\frac{7}{4r_{0}^{2}}-\frac{7a_{5}}{8}\right)\left(\frac{r_{0}}{r}\right)^{4}+a_{5}\left(\frac{r_{0}}{r}\right)^{5}
\end{align*}

 \begin{figure}
\centering
\includegraphics[width=0.4\textwidth]{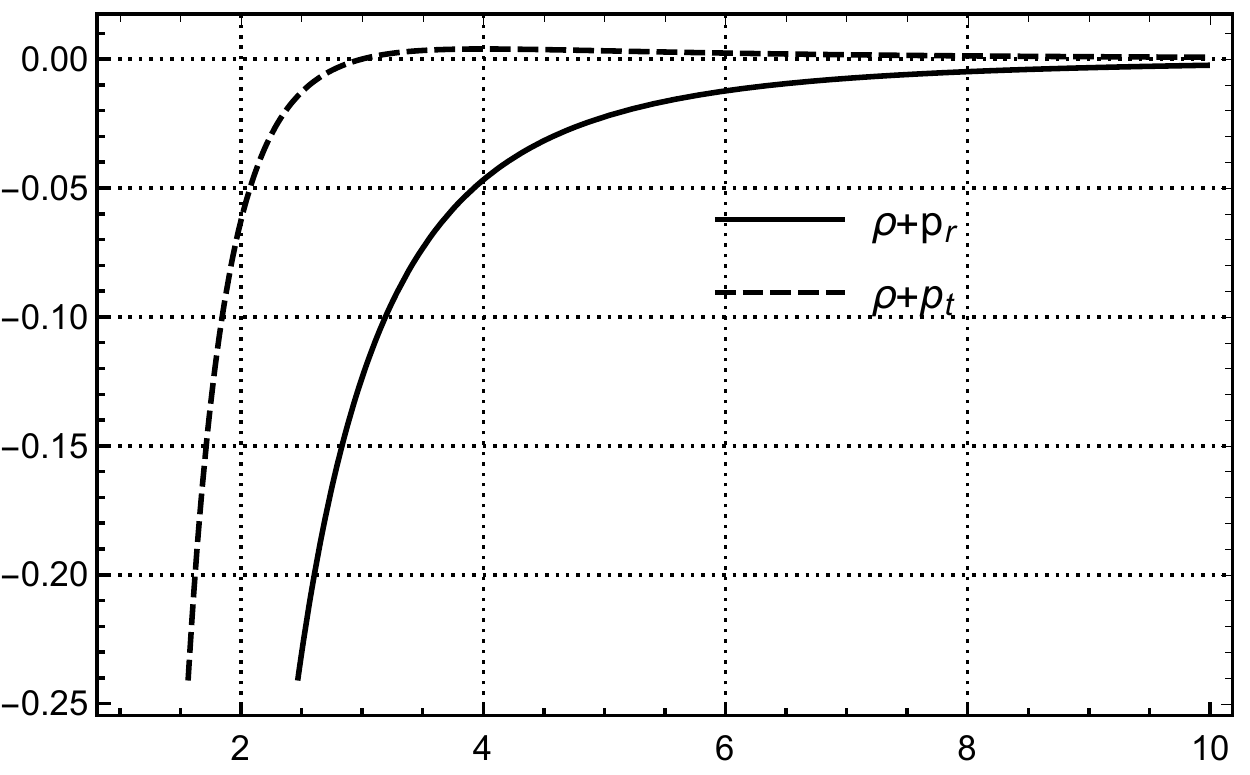}\qquad
\includegraphics[width=0.4\textwidth]{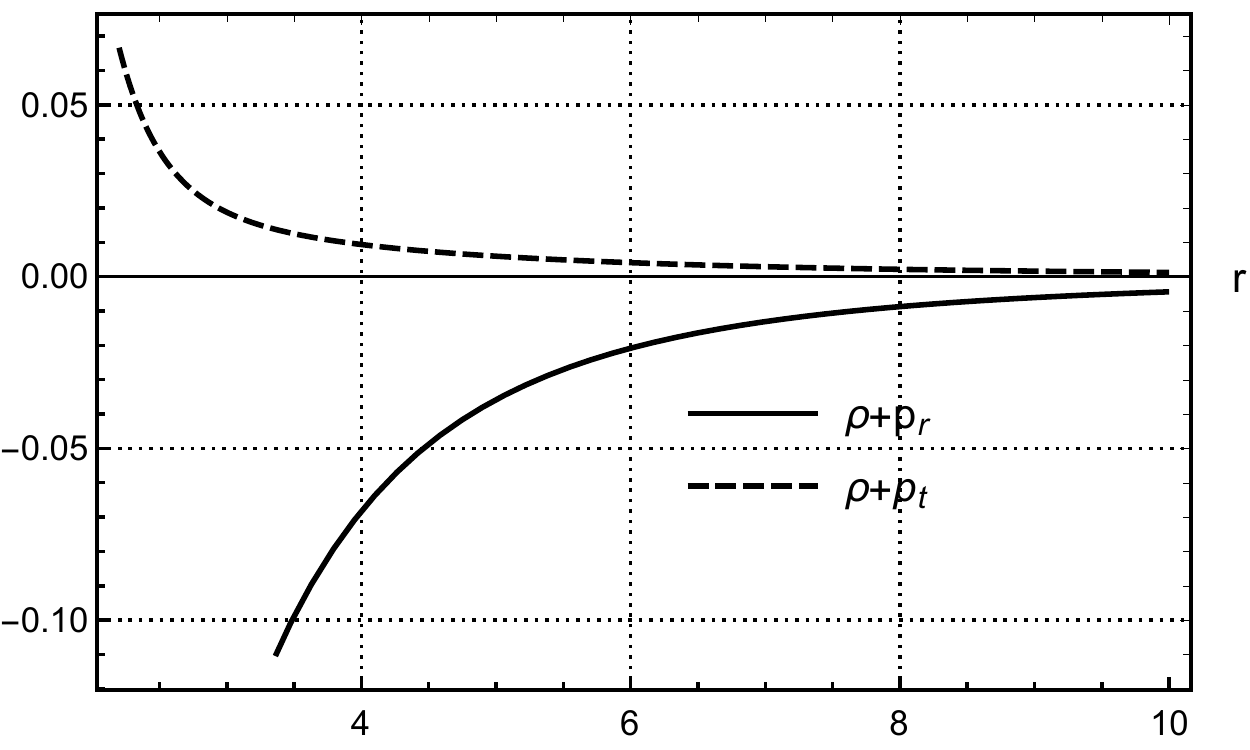}

\caption{The energy conditions for wormholes with $n=4$ and $n=5$ respectively. }
\end{figure}

\begin{figure}
\centering
\includegraphics[width=0.4\textwidth]{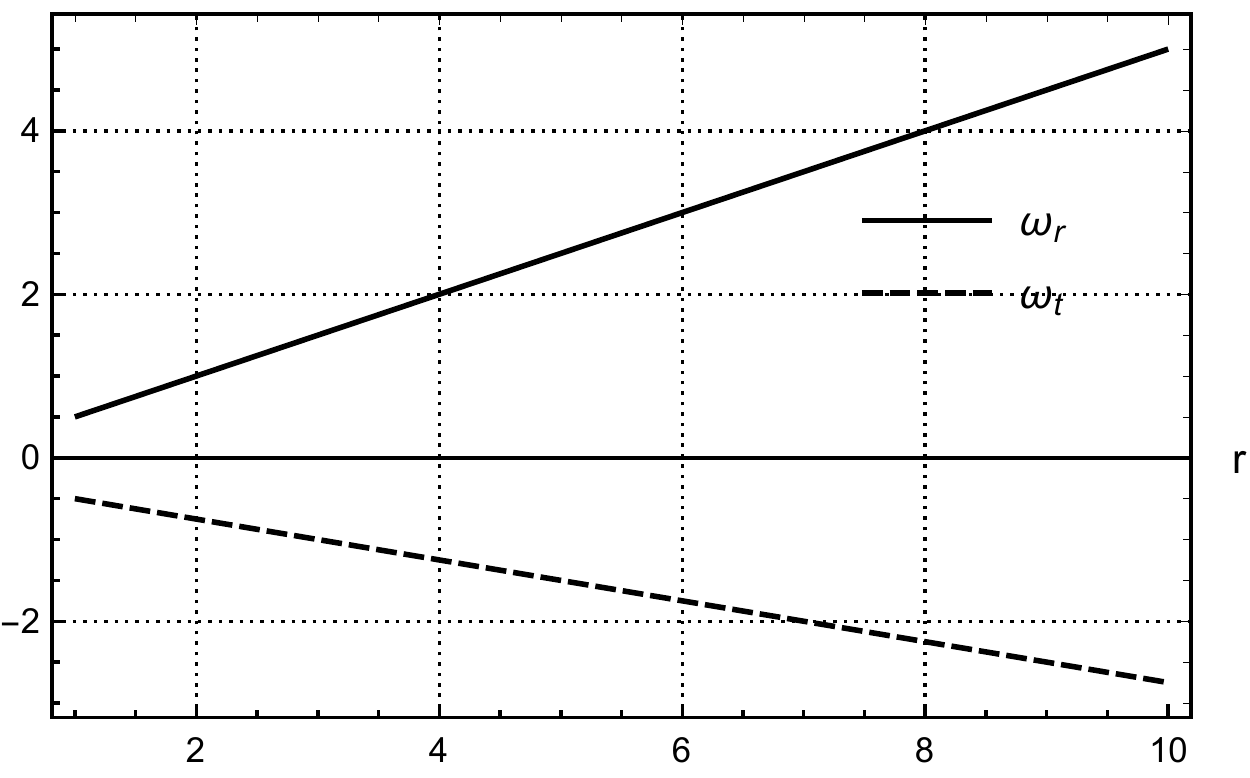}\qquad
\includegraphics[width=0.4\textwidth]{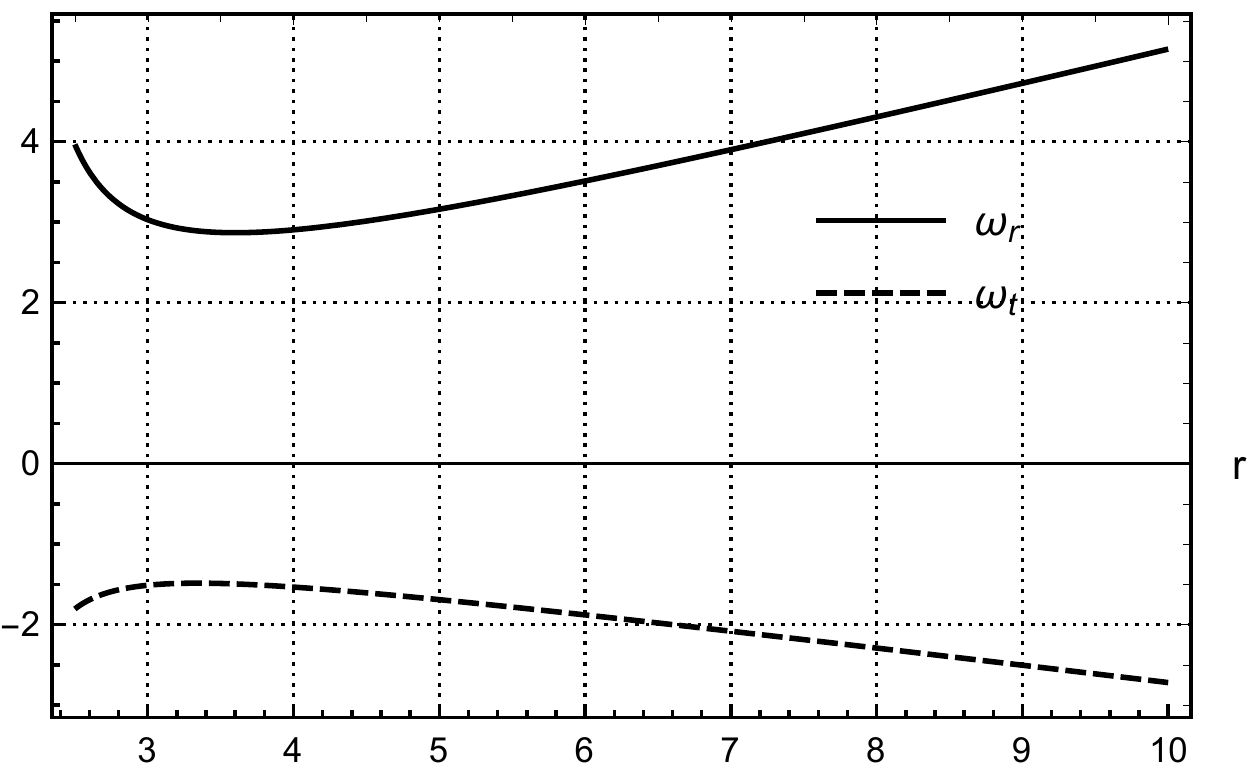}
\caption{The equation of state parameters for wormholes represented using $n=4$ with $r_{0}=2$ and $n=5$ with $r_{0}=3$ respectively. }
\end{figure}

\subsubsection{The Energy Conditions}

Since we are constructing traversable wormholes, by construction the supported matter tensors near the throat violated null energy conditions (NEC) as is seen in Fig.1. Further the above class of wormholes constructed by the polynomial complexity function and logarithmic red-shift factor, can be classified as wormholes of Casimir class. Casimir energy density is an example of exotic matter that can be obtained in a laboratory and is presumed significant in the context of traversable wormholes. In \cite{casi1} wormholes supported by Casimir energy were constructed, which came to be known as the Casimir wormhole. The Casimir energy density follows the equation $p_{rc}=\omega_{rc}\rho_{c}$ and $p_{tc}=\omega_{tc}\rho_{c}$ where $(\rho_{c},p_{rc},p_{tc})$ are the corresponding Casimir energy density, radial pressure and tangential pressure respectively. In case of Casimir the equation of state $\omega_{rc}$ for $p_{rc}$ is typically 3, with $\rho_{c}$ exotic. Wormholes supported by Casimir energy are treated in a variety of context in the literature \cite{5, casi}. The wormhole constructed for $n=5$ falls exactly into the category of a Casimir wormhole with the constraint $-2<a_{5}r_{0}^{2}\leq 0.$ This can be verified from the figure 2 where $\omega_{r}=\frac{p_{r}}{\rho}$ which is nearly 3 around the throat evaluated at $r_{0}=3.$ The wormhole constructed for $n=4$ is a Casimir like wormhole with $\omega_{r}$ positive for all values of $r.$ Fig.2 gives a graphical representation for $\omega_{r}$ and $\omega_{t}=\frac{p_{t}}{\rho}$ in the wormhole examples for $n=4$ and $n=5.$

{\section{Wormholes with ``zero complexity"}

It has been already noted that for wormholes only way by which a zero complexity can be achieved is when $(p_{r}-p_{t})=\frac{1}{2r^{3}}\int_{0}^{r}s^{3}\rho'(s)ds.$ We will now explore this scenario to comment on the viability of such existence. From equation (\ref{me}) and $y_{tf}=0$ we get 
\begin{equation}
\frac{b(r)}{r}=1-\left(\frac{Ce^{2\phi}}{(r^{4}f'(r))^{2}}\right)f(r)\label{shpz}
\end{equation}
where $f(r)=\int\frac{e^{2\phi}\phi'}{r^{5}}dr$ is a continuous and differentiable function of $r$ and $r^{5}f'(r)=e^{2\phi}\phi'$ with `prime' denoting differentiation w.r.t $r.$ Considering the restrictions on viable traversable wormhole we know that for $r\rightarrow\infty,~\phi'\rightarrow 0,~e^{2\phi}\rightarrow$ `a finite limit'. This gives $\lim\limits_{r\to\infty}(r^{5}f'(r))=0.$ We now claim that \emph{``General zero-complexity traversable wormholes are not feasible".} 
Accordingly we provide the following proof.
\begin{proof}
For (\ref{shpz}) to represent a viable wormhole throat we must have 
\begin{itemize}
\item $\frac{b(r)}{r}\rightarrow 0$ as $r\rightarrow\infty,$ this gives $\frac{e^{2\phi}}{(r^{4}f'(r))^{2}}f(r)\rightarrow\frac{1}{C}$ as $r\rightarrow\infty.$
\item At the throat $r_{0}=b(r_{0})$ which gives $f(r_{0})=0$ since $\frac{e^{2\phi(r_{0})}}{(r_{0}^{4}f'(r_{0}))^{2}}\neq 0.$
\end{itemize}
By the above we have $f(r_{0})= 0$ and $f(r)\rightarrow 0$ as $r\rightarrow\infty.$ This gives that there is at least one $r=r_{1}\in(r_{0},\infty),$ such that $f'(r_{1})=0.$ 

The above is possible as follows: 
\begin{description}
\item Let if possible $f'(r)>0$ for $r\geq r_{0}.$ Then $f(r)$ is a monotonic increasing function on $[r_{0},\infty)\Rightarrow f(r)\nrightarrow 0$ as $r\rightarrow\infty.$ A contradiction.
\item Let if possible  $f'(r)<0$ for $r\geq r_{0}.$ Then $f(r)$ is a monotonic decreasing function on $[r_{0},\infty)\Rightarrow f(r)\nrightarrow 0$ as $r\rightarrow\infty.$ A contradiction.
\item Finally let $f'(r)>0$ for some interval $I\subset[r_{0},\infty)$ and let $f'(r)<0$ for some interval $J\subset[r_{0},\infty).$ Since $f'(r)$ is a continuous function, we must have $f'(r)=0$ for some $r=r_{1}\in (r_{0},\infty).$ This gives $f'(r)=0$ for at least one $r\in[r_{0},\infty).$
\end{description}

The zero in $f'(r)$ is expressed as a singularity in $b(r)$ for $r\in[r_{0},\infty).$ Thus we see that viable zero-complexity traversable wormholes are in general not possible. 
\end{proof}
 This shows that traversable wormholes can never be ``simple".  Wormholes in general have complex structure, with the complexity in them arising due to the traversable conditions imposed. In order to make the geometry suitable for traversal the matter stress tensors become exotic violating the NEC. Interestingly we see that this fact is represented through the definition of the recently proposed complexity factor. Although complexity factor can be zero due to couple of factors, (one being physical and the other being mathematical) in case of wormholes as is expected, physical conditions prevents imposition of mathematical simplicity to the traversable wormhole which shows that they can never be equated to the class of zero-complexity spherically symmetric gravitational objects.

\section{Conclusion}

We have considered a complexity factor parametrization of the traversable wormhole. Using the recently introduced complexity factor definition we have found viable traversable wormhole geometry. We proposed a general method of defining polynomial complexity factor based on which we have proposed analytical methods by which we can obtain any desired traversable wormhole configurations. Further our results show that traversable wormholes can not be of zero-complexity variety. Imposing the restrictions of asymptotic flare-out and no horizon prevents wormhole geometry with $y_{tf}=0.$ This is the only natural possibility, since complexity of a self gravitating system includes the contributions of matter density inhomogeneity and pressure anisotropy, which in the case of a traversable wormhole is always inhomogeneous and usually anisotropic. The non-zero complexity factor parametrized wormholes so obtained could be classified as the Casimir wormholes \cite{5, casi1, casi}. The choice of the logarithmic tidal function and polynomial complexity factor makes them Casimir like wormholes with positive equation of state $\omega_{r}$ corresponding to the radial pressure density. It may be noted that for $n=5$ we could obtain a Casimir wormhole and the supporting matter in the neighbourhood of the throat having $\omega_{r}\simeq 3.$

\section{Acknowledgments}
 SB acknowledges IUCAA, Pune, India, for their warm hospitality while working on this project.

\end{document}